# Graphical Abstract

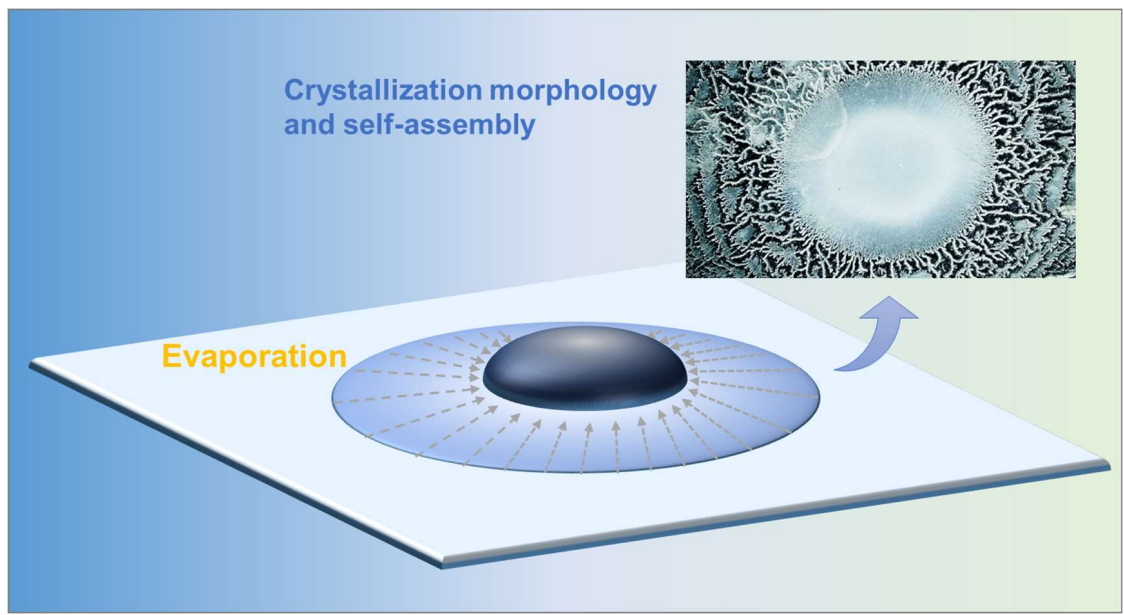

# Crystallization morphology and self-assembly of polyacrylamide solutions during evaporation

Jun Hu[1,2], Zhan-Long Wang[1,*]

[1]Shenzhen Institute of Advanced Technology, Chinese Academy of Sciences, Shenzhen, Guangdong 518000, China.

[2]China University of Political Science and Law, Beijing, 100091, China.

E-mail: zl.wang1@siat.ac.cn

**Abstract**

This study investigates the crystallization and self-assembly phenomena of polyacrylamide (PAM) solutions during evaporation. While traditional thin-film fabrication methods such as spin coating and drop casting are commonly used, this study utilizes a simple evaporation approach to gain insights into the self-assembly processes of PAM solutions. We examined PAM solutions with varying concentrations (1%, 2%, and 5%) and observed the resulting crystalline structures and morphological changes during evaporation. Spectroscopic absorbance measurements were employed to analyze concentration gradients and crystallization patterns. The findings reveal that the concentration gradient during evaporation significantly impacts the crystallization behavior of PAM solutions. The 2% solution exhibited the clearest and most regular crystalline patterns, whereas the 1% and 5% solutions displayed more complex and irregular morphologies. These observations provide new insights into how solution concentration influences PAM crystallization, contributing to a better understanding of polymer thin-film self-assembly. The novelty of this study lies in exploring PAM self-assembly through controlled evaporation, offering valuable insights for the development of polymer films.

# 1. Introduction

Polymer films have gained significant attention because of their wide range of applications in various industries such as electronics, membranes, coatings, and sensors[1-5]. The crystallization, self-assembly, and resulting patterning of these films play crucial roles in determining their properties and performance[6-11]. Evaporation of a polymer solution is an important method for fabricating polymer films[12-15]. Generally, by controlling the concentration of polymer solutions, volatility of solvents, evaporation rate, and other parameters, the deposition of polymers and formation of thin films can be effectively achieved[12,13,16-19]. Examples include the preparation of self-assembled and coated films. In the preparation of self-assembled thin films, polymers are mixed with solvents to form a solution, and the concentration of the polymers in the solution is controlled through solvent evaporation to allow self-assembly into thin films[20,21]. The solution evaporation method is commonly used to prepare coated films. In this method, a polymer solution is coated onto the substrate, and the polymer is then deposited and formed into a thin film through solvent evaporation[22]. This method is also used to prepare transparent conductive films such as zinc oxide or polymer conductive films[23,24]. Electrospinning is a method of spinning polymer solutions into fibers using electrostatic forces, and the evaporation of the solution is a key step in the spinning process. After the polymer solution passes through the spinning nozzle, the solvent evaporates, and the polymer is deposited to form a thin fiber film[25]. Films prepared by solution drop coating are used to prepare organic optoelectronic devices, sensors, and thin-film substrates. Therefore, the evaporation of polymer solutions is important in many cutting-edge research fields.

In the past, extensive research has been conducted in the field of thin film formation by polymer solution evaporation. Researchers have explored various methods such as spin coating, drop casting, and immersion to achieve uniform evaporation and thin-film formation of polymer solutions[22,26-30]. The influence of various solution parameters, including the polymer concentration, evaporation rate, and characteristics of additives and solvents, on the morphology and properties of the films has been investigated [12,13,16-19]. Significant progress has been made in studying the structure, thickness, and surface morphology as well as mechanical, optical, and electrical properties of films and exploring ways to control the film structure and properties through polymer solution evaporation[31,32]. Polymer films formed by solution evaporation have been applied in diverse fields, including optoelectronic devices (such as solar cells), sensors, flexible electronics, and biomedicine[33,34]. Fundamental research on these mechanisms involves studying the behavior of liquid droplets, interfacial phenomena, and mass transfer during polymer solution evaporation, with the aim of gaining a deeper understanding of the underlying principles of thin film formation[18,35-45]. These research areas focus on understanding the mechanisms that control the morphology, tune the properties, and expand the applications of polymer solution-evaporated films, thereby providing a foundation for the development of this technology. However, despite significant progress in understanding the mechanisms, applications, performance, and evaporation patterns of polymer solution-evaporated films, numerous phenomena and issues remain unresolved in this field.

Polyacrylamide (PAM) is an important synthetic polymer with a high molecular weight, high water absorption, excellent stability, ecological friendliness, and other characteristics. Nowadays, PAM has been widely used in water treatment, thickeners, colloid stabilizers, oil mining, textile and paper industries, hydrogel materials and their applications, and many other fields[46,47]. PAM also exhibits good water absorption and can rapidly form gels in water. Its high molecular weight and water absorption properties make it a suitable candidate for investigating how concentration gradients and internal flow fields drive crystallization during solvent evaporation. However, the crystallization of PAM solutions, and the morphology formed after the evaporation of water are still been understood thoroughly.

In this study, we studied the different crystal morphologies and self-assembly phenomena formed by the evaporation of the PAM solution thin film formation. We conducted experiments using PAM solutions at concentrations of 1%, 2%, and 5% and examined their drying behavior in transparent plastic culture dishes. Through visual observation, we characterized the crystallization patterns that emerged after solvent evaporation. The crystallization of the PAM solution showed three stages: a chaotic particle pattern from the center to the edge, a mountain pattern, and a grass-like pattern. This phenomenon can be attributed to the induction of concentration changes generated by internal flow fields during evaporation. The absorption spectrum indicated that during the evaporation of the solution, the non-dried portion gradually increased with the evaporation concentration. This work reveals a new phenomenon of polymer solution crystallization and self-assembly, which may expand our understanding of polymer solution crystallization and provide guidance for thin film preparation.

## 2. Materials and methods

The solutions were prepared by mixing PAM powder and deionized (DI) water at mass ratios of 1:99,

1:49, and 1:19, corresponding to mass percentages of 1%, 2%, and 5%, respectively. The PAM powder was purchased from Shanghai Aladdin Company, with a molecular weight range from 4000 to 40000. To ensure consistency and accuracy in solution preparation, specific precautions were taken: all glassware used was pre-cleaned with DI water, and the PAM powder was stored in a desiccator to prevent moisture absorption before use. For each solution preparation, the exact volume of DI water was measured and added to a 100 ml beaker. The PAM powder was then gradually introduced while stirring to avoid agglomeration. The stirring process was carried out using a magnetic stirrer (IKA RCT basic) with a stirring speed set to 800 rpm. This speed was selected based on preliminary optimization experiments, where different speeds were tested to balance between rapid dissolution and avoiding foam formation or vortex generation. A stirring speed of 800 rpm was found to be optimal for ensuring the complete dissolution of the PAM powder within 2 hours, producing a homogeneous solution. This dissolution process was monitored carefully to avoid powder aggregation. The clarity of the prepared solutions was inspected visually and using optical microscopy. Specifically, after stirring, the solutions were allowed to stand for a few minutes, and the absence of visible suspended particles or lumps was confirmed under a microscope at 10x magnification. This method ensured that the solutions were homogeneous and free of undissolved polymer particles before further use. After preparation, the viscosity of the PAM solutions was noted to increase with higher concentrations of PAM powder. The prepared solutions were shown in Fig. 1b. To study the crystallization behavior, droplets of each PAM solution were deposited on sterile culture dishes using a calibrated micropipette (Eppendorf Research plus) to ensure uniform droplet volumes of 5 μl. The culture dishes were placed in a controlled environment chamber at 25°C with a relative humidity of 60%, where the solution was allowed to evaporate spontaneously. The volume of each solution was reduced to 5 ml before droplet deposition to ensure comparable evaporation times across all concentrations. The evaporation process and subsequent formation of crystalline patterns were monitored periodically using optical microscopy. The droplets were observed every 30 minutes until full evaporation occurred. The total evaporation time ranged from 12 to 36 hours, depending on the concentration of the PAM solution. The resulting crystalline structures were photographed and analyzed to determine the influence of concentration on crystal morphology.

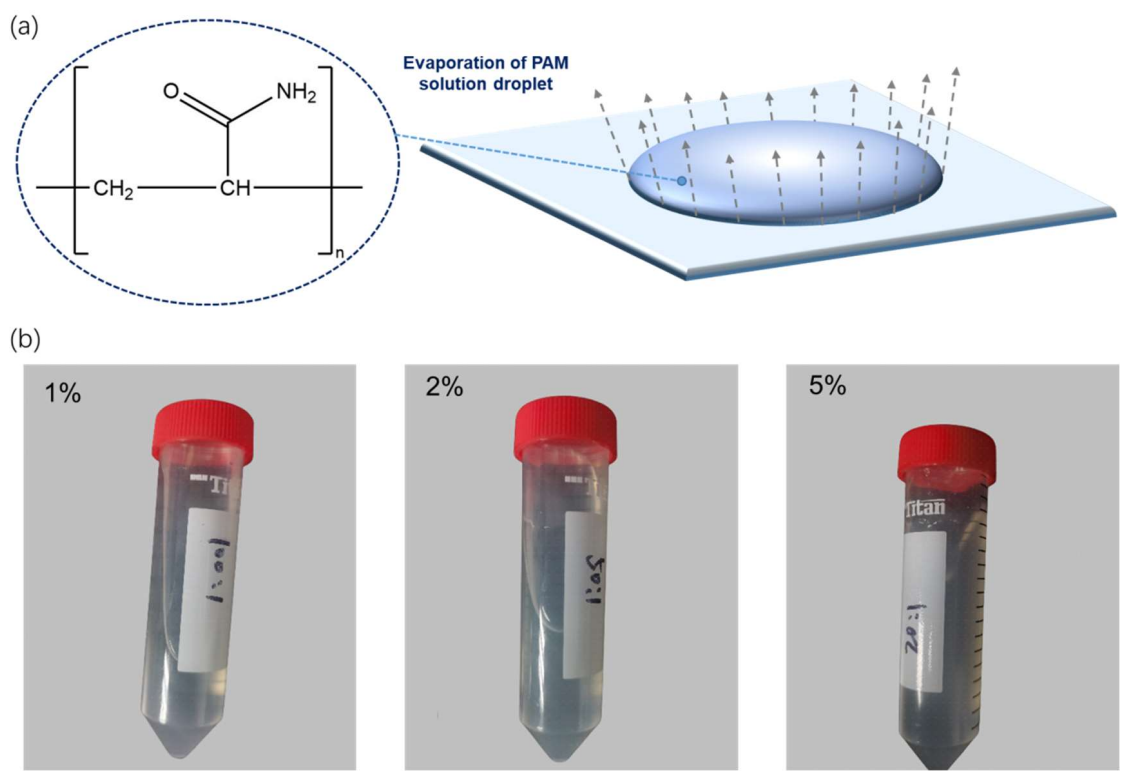

**Figure 1**. (a) The diagram of PAM solution droplet evaporation; (b) the prepared mixed PAM solutions.

## 3. Results and discussion

### 3.1 *The macroscale observation of PAM solutions evaporation*

In Fig. 2, the drying behavior of PAM solutions at 1%, 2%, and 5% concentrations in transparent plastic Petri dishes with a diameter of 3.5 mm is presented. Figure 2a–c represent the 1%, 2%, and 5% concentrations, respectively. For the 1% solution in Fig. 2a, as the solvent evaporated, the PAM solution formed a nonuniform thin film. This crystalline morphology suggests that the polymer molecules in the 1% PAM solution arranged in an ordered manner during drying, resulting in a tightly packed structure. In the case of the 2% solution, as shown in Fig. 2b, distinct structural differences were observed among the three different regions of the dried crystalline morphology. This indicates that significant variations in the crystallization process occurred within different regions of the 2% solution. These local morphological differences may be attributed to concentration gradients and differences in the arrangement of the polymer molecules in different concentration regions[48]. In contrast, the 5% solution exhibited two distinct crystalline morphologies upon drying, as shown in Fig. 2c. These differences can be attributed to local concentration gradients that form during the evaporation process. As the solvent evaporates, the polymer concentration increases unevenly across the Petri dish due to variations in evaporation rates and polymer diffusion. At higher concentrations, such as in the center, the PAM molecules have limited mobility, leading to the formation of tightly packed, highly ordered structures. In contrast, at the periphery, where the concentration is relatively lower and evaporation occurs more rapidly, the polymer chains have more mobility, resulting in a looser, more disordered crystalline structure. This could be due to the formation of local concentration gradients under high-concentration conditions, leading to variations in the crystallization behavior in different regions. Consequently, two different crystalline morphologies were formed. Among these observations, the 2% solution displayed the most distinct and regular crystalline pattern. This may be attributed to the fact that at moderate concentrations, the arrangement of polymer molecules is better controlled, resulting in the formation of a more ordered structure during crystallization[49]. In contrast, both the 1% and 5% solutions showed less defined crystalline patterns, likely due to the challenges of initiating crystallization at low and high concentrations. At low concentrations, molecular interactions may be too weak to drive crystallization, while at high concentrations, increased viscosity may hinder molecular mobility, preventing uniform structure formation.

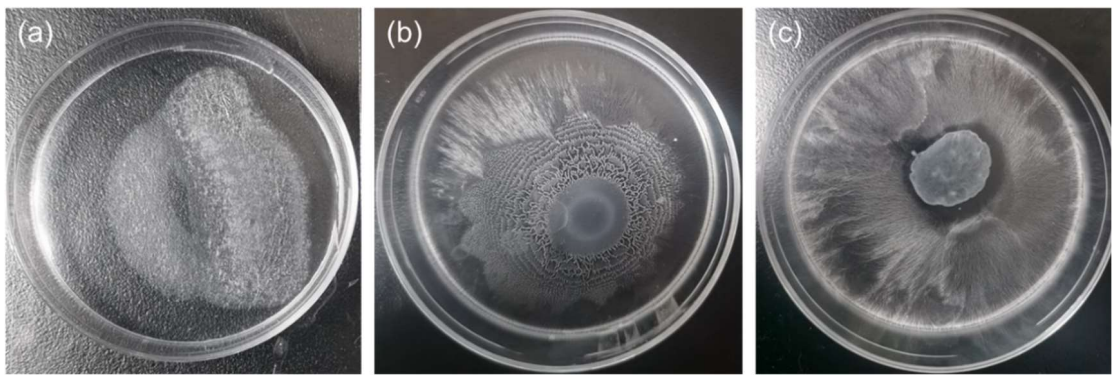

**Figure 2**. Images of the patterning during PAM solution evaporation in plastic culture dishes. (a)–(c) show the crystalline morphology formed by drying 1%, 2%, and 5% PAM solutions in culture dishes, respectively.

### 3.2 *The microscale observation of crystallization and self-assembly*

Evaporation-induced crystallization of the 2% PAM solution in plastic Petri dishes resulted in the formation of intricate crystal morphologies, as shown in Fig. 3. Figure 3a illustrates three distinct morphologies that expand outward from the center. At the center, disordered granular crystals were observed, representing the first level of crystallization. Surrounding the granular crystals, in the second level of crystallization, a network resembling mountain ridges emerged, with intercalated petal-like structures. Furthermore, in the third level of crystallization, independent grass-like crystal textures were formed, exhibiting branch-like or feather-like morphologies. Figure 3 illustrates this phenomenon, depicting the macroscopic and microscopic morphologies. Figure 3a shows macroscopic images of the three morphological levels, while Figures 3b to 3f provide microscopic views of each level and their transition regions. These hierarchical structures arise from a combination of polymer-solvent interactions and solvent evaporation dynamics. The kinetics of solvent evaporation and the resulting concentration gradients play a significant role in determining the spatial variation of crystal formation. Evaporation rates were nonuniform across the dish surface, influencing polymer mobility and leading to the hierarchical self-assembly process observed.

Figure 3 shows the change in the crystal morphology during the evaporation of the PAM solution. This complex crystallization behavior can be attributed to several factors. The concentration of the PAM solution plays a critical role in determining the resulting morphology. This can be explained by variations in local polymer concentrations and differential rates of solvent evaporation. As the solvent evaporates, the polymer molecules undergo conformational changes to minimize free energy, facilitating the self-assembly of ordered structures. The formation of granular clusters, ridge-like networks, and grass-like structures is governed by these solvent evaporation dynamics, which have been previously demonstrated to drive similar hierarchical structures in polymer systems. The 2% concentration used in this study favored the formation of hierarchical structures owing to the interplay of polymer-solvent interactions and solvent evaporation kinetics. As the solvent evaporated, the polymer molecules underwent self-assembly, leading to the observed hierarchical organization. In addition, the kinetics of solvent evaporation contributed significantly to the formation of diverse morphologies[50]. The spatial variation in evaporation rates across the Petri dish surface led to differential concentration gradients, influencing the nucleation and growth of crystals. This results in the formation of distinct morphological features such as granular clusters, ridge-like networks, and grass structures. Furthermore, the interplay between the polymer chain mobility and solvent diffusion during evaporation governs the hierarchical assembly process. As solvent molecules escape, the polymer chains undergo conformational changes and aggregate to minimize free energy, giving rise to the observed multilevel morphologies[51]. Among the numerous factors involved, solvent evaporation dynamics have emerged as a pivotal aspect shaping the complex crystallization behavior. Solvent evaporation dynamics encompass the interplay between the various processes that occur during solvent removal, such as diffusion, concentration gradients, and surface effects.

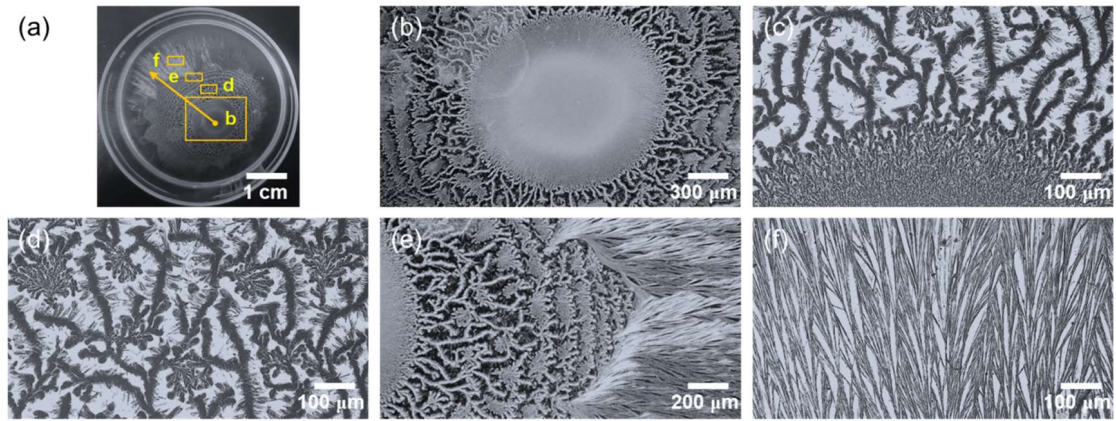

**Figure 3**. The microscopic images of the crystal morphology of PAM solutions in a culture dish. (a) The whole and macroscopic view of the solution crystal morphology. The crystal presents a three-stage gradient change. (b) The microscopic crystal morphology in the first stage in the center of culture dish. (c) The microscopic view of the crystal at the junction of the first and second stages. (d) The microscopic crystal morphology in the second stage. (e) The microscopic view of the crystal at the junction of the second and third stages. (f) The microscopic crystal morphology in the third stage.

At the heart of solvent evaporation dynamics is the concentration gradient established as solvent molecules escape from the solution into the surrounding environment. This concentration gradient induces differential rates of solvent removal across the solution, leading to localized variations in the polymer concentration. These concentration variations influence the nucleation and growth processes, ultimately shaping the morphology of the formed crystalline structures. Moreover, the spatial and temporal variations in the solvent evaporation rates contribute to the formation of distinct regions within the crystallizing solution, each characterized by unique concentration profiles. These concentration gradients act as driving forces for the migration of the polymer chains, leading to the assembly of ordered structures at multiple length scales. The observed hierarchical organization, such as the formation of particle clusters, mountain-like networks, and branching structures, can be attributed to spatially heterogeneous solvent evaporation dynamics. Furthermore, the dynamics of the polymer chain motion are intricately linked to solvent evaporation. As solvent molecules escape, the polymer chains undergo conformational changes and self-assembly to minimize free energy. Competition between chain mobility and polymer–polymer interactions underlies the emergence of hierarchical structures during crystallization. Surface effects also occur as solvent molecules evaporate more rapidly from the exposed surfaces of the solution, leading to enhanced polymer concentration gradients and localized crystallization. The confinement imposed by the container surface further influences the spatial distribution of the polymer chains, contributing to the formation of structured morphologies. Solvent evaporation dynamics have emerged as critical factors governing the complex crystallization behavior observed in PAM solutions. The establishment of concentration gradients, differential rates of solvent removal, polymer chain dynamics, and surface effects collectively drive the hierarchical assembly of the crystalline structures. Understanding and manipulating these dynamics offers avenues for tailoring the properties of crystalline materials across various length scales, with implications for applications in materials science, biotechnology, and beyond.

### 3.3 *Detailed observation of local sections*

In Fig. 4, we present the results of crystallization observed in the third stage, which is characterized by the formation of dendritic or grass-like structures within the PAM solution. These dendritic or grass-like crystals exhibit a characteristic morphology, with individual branches possessing widths of approximately 10 μm and lengths ranging from 500 μm to 1 mm. The overall appearance is a grass-like crystalline structure. Figure 4 shows micrographs of different regions of these crystals, highlighting their observed characteristics. Figure 4a shows an overview of the overall morphology of the grass-like crystals formed by the crystallization of the PAM solutions. Figure 4b-d show details of the crystal morphology at different locations in the complete grass-like crystal. The formation of the dendritic and grass-like structures can be attributed to several factors. One of the contributing factors is the nucleation and growth kinetics of the solution. As the polymer solution cools or evaporates, localized regions with higher polymer concentrations can initiate the formation of nucleation sites. These sites serve as points for crystal growth, with polymer molecules aggregating and aligning along specific directions dictated by the intermolecular forces and chain conformations. The dendritic morphology arises from the preferential growth along certain crystallographic axes, resulting in branching structures. Furthermore, the presence of impurities or additives in the solution can influence the crystallization process. Impurities may act as heterogeneous nucleation sites, promoting the formation of crystals with nonuniform shapes and branching patterns. In addition, the interactions between the polymer and solvent molecules play a crucial role in determining the final morphology of the crystals. Variations in the solvent composition, temperature, and concentration gradients can affect the crystallization process, leading to the observed diversity in crystal morphology.

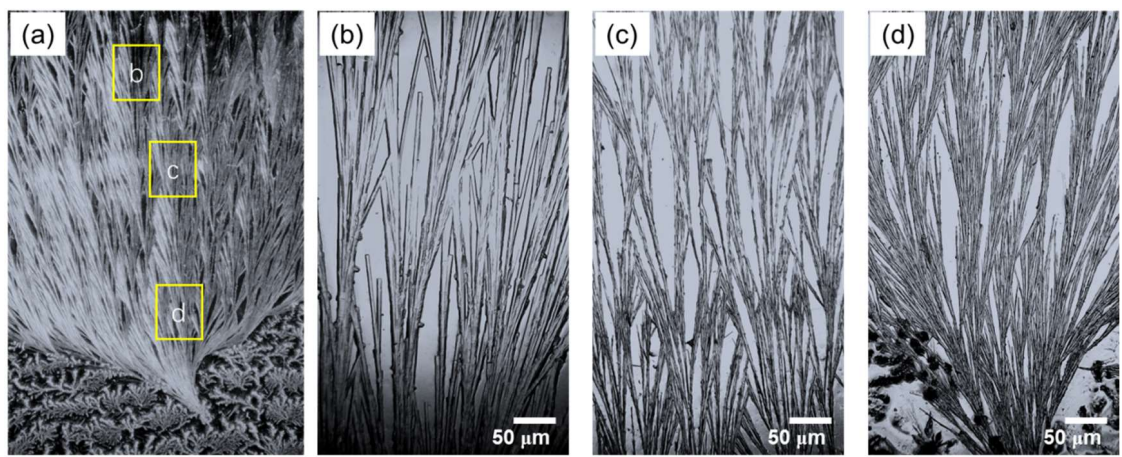

**Figure 4**. The tussock-like crystals formed in the third-level crystallization area. (a) The enlarged image of the grass-like crystal structures in the third level. (b) The microscopic view of the dendritic or grass-like structures in the tail section of the crystal. (c) The microscopic images in the middle section of the grass-like crystal structures. (d) The microscopic view of the grass-like crystal in the section near the crystal root.

In Fig. 5, we observe the crystallization patterns in the second stage, which are characterized by the formation of mountainous vein-like structures. These crystals exhibit irregular vein textures with widths of approximately 10 μm, and gaps exist between the veins where petal-shaped crystals are generated. Micrographs of different regions of these crystals clearly show these observed characteristics (Fig. 5a–f). The formation of these mountainous vein-like structures in the PAM solution can be attributed to multiple factors. The interplay among the polymer concentration, solvent evaporation rate, and temperature gradient within the solution is a key factor. As the solvent evaporates or the solution cools, the polymer molecules tend to aggregate and align along specific directions, leading to the formation of vein-like structures. Irregularities in the vein texture and the generation of petal-shaped crystals in the gaps between veins may be influenced by variations in the local polymer concentration, solvent composition, and presence of impurities. In addition, crystal growth dynamics play a crucial role in shaping the observed morphology. The growth of petal-shaped crystals in the gaps between the veins may have resulted from the preferential attachment of polymer molecules to these sites, leading to the formation of distinctive features. Furthermore, irregular vein textures could arise from variations in the growth rates along different directions, indicating a complex interplay of kinetic and thermodynamic factors during crystallization.

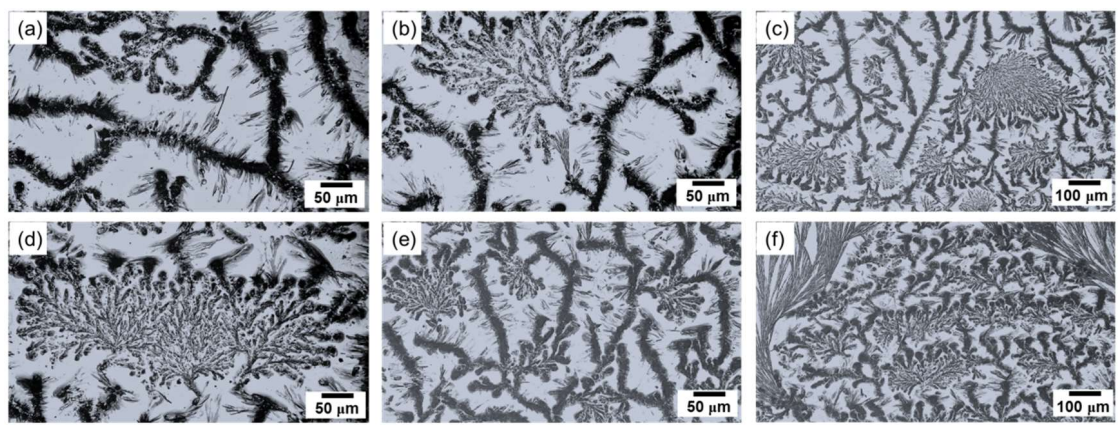

**Figure 5**. The veined crystals forming in the second-level crystallization area. (a)–(f) Different views of the veined crystals.

Figure 6 sheds light on the crystallization phenomena observed in Figs. 3 to 5 and delves into their underlying mechanisms. Absorbance spectroscopy analysis reveals a gradual increase in solution concentration during the PAM solution evaporation process. This concentration gradient, with lower concentrations at the periphery and higher concentrations towards the center, contributes to the formation of the tri-level structure depicted in Fig. 3. Figure 6a to 6f depict schematic diagrams of evaporation stages and absorbance spectra. Figure 6a corresponds to the initial evaporation stage, showcasing the formation of the third-level grass-like crystalline structure observed in Fig. 3. Figure 6b illustrates an intermediate stage of evaporative crystallization, where increasing solution concentration leads to the formation of the second-level mountainous vein-like crystalline structure. Ultimately, a further increase in solution concentration results in the formation of the first-level irregular evaporative crystal morphology. The absorbance spectra depicted in Figs. 6d, 6e, and 6f exhibit a consistent trend of increasing absorbance intensity, indicating a gradual rise in solution concentration. This observation aligns with the presence of concentration gradients observed during the evaporation process, which play a pivotal role in shaping the intricate crystalline structures depicted in Figs. 3 to 5. Through this systematic examination of the evaporation process and its correlation with solution concentration, valuable insights are gained into the mechanisms dictating the formation of complex crystalline morphologies within the PAM solution. By elucidating the relationship between concentration gradients and crystallization behavior, this analysis deepens our understanding of the intricate interplay between thermodynamic and kinetic factors governing polymer crystallization processes.

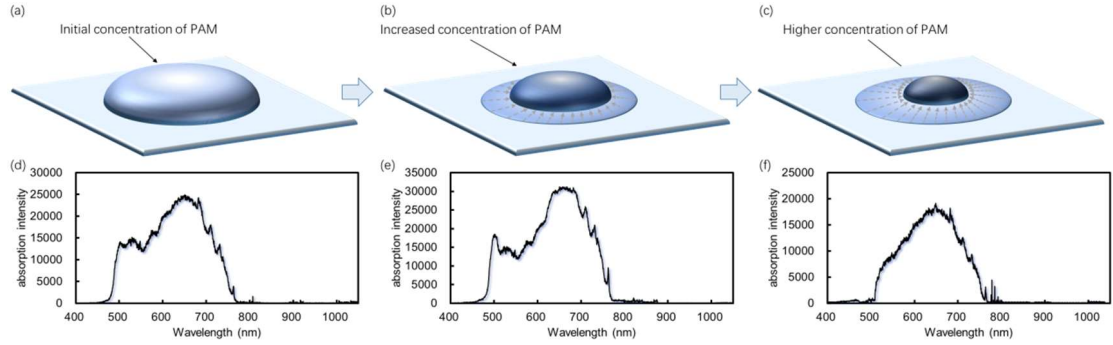

**Figure 6**. Diagram of the mechanism of PAM solution evaporation on a solid surface. (a)–(c) Diagram of the progress of PAM solution evaporation. As the PAM solution evaporates, the concentration of the residual solution increases under the internal flow fields. (d)–(f) The absorption spectrograms of the solutions in different stages. The spectrograms show that the absorption intensity gradually increases as the PAM solution evaporates.

The observed increase in absorbance intensity followed by a subsequent decrease during the evaporation process can be explained by the dynamic changes in polymer concentration and structural evolution in the thin films. Initially, as solvent begins to evaporate from the PAM solution, the polymer concentration in the solution increases. This increase in concentration leads to greater interaction between polymer chains, causing them to aggregate and form denser structures. As the polymer chains aggregate, light scattering within the solution increases, leading to a rise in absorbance intensity. This is indicative of the growing density of the polymer network, as more polymer chains come into close proximity, causing the solution to become increasingly turbid. The initial rise in absorbance is a direct result of the increased concentration of polymer molecules in the evaporating solution, which enhances the absorption of light. Additionally, the formation of small crystalline or semi-crystalline domains within the film may contribute to the higher absorbance, as these structures have different refractive indices compared to the surrounding medium. The growth of these crystalline structures, along with increasing polymer concentration, amplifies the overall absorbance. As the solvent content decreases, the alignment and self-assembly of the polymer chains become more pronounced, further increasing the absorbance due to the organized nature of the crystalline regions. However, after reaching a peak absorbance, the intensity begins to decrease, which can be attributed to several factors. One of the primary reasons for the drop in absorbance is the depletion of solvent, which leads to the cessation of molecular mobility in the polymer solution. At this stage, the polymer chains may have reached a point of maximum packing density, and as evaporation continues, the film becomes increasingly solidified. This transition from a liquid-like or semi-crystalline state to a more rigid, solidified film reduces the ability of the system to absorb light, as the polymer chains are no longer free to rearrange or interact with incoming light in the same way they did during the earlier stages of evaporation. Furthermore, as crystallization completes and the film dries, scattering effects may diminish, resulting in lower absorbance. In addition, the formation of larger crystalline domains during the latter stages of evaporation may lead to phase separation or structural regularization, which could reduce the scattering of light and thus lower the overall absorbance. This behavior is consistent with the transition from a disordered, amorphous phase to a more ordered, crystalline phase, where the refractive index contrast between the polymer and solvent is reduced, thereby lowering absorbance. Another potential explanation for the decrease in absorbance is the emergence of surface tension effects as the film solidifies. As the solvent evaporates and the film becomes more rigid, surface tension forces may act to smooth out the surface of the film, reducing the roughness and thereby decreasing the scattering of light. This reduction in scattering would further contribute to the observed decrease in absorbance. Thus, the initial increase in absorbance intensity is driven by the concentration of polymer molecules and the formation of dense, semi-crystalline structures during evaporation. The subsequent decrease in absorbance can be attributed to the completion of the crystallization process, reduced light scattering as the film solidifies, and the structural organization of the polymer chains into more ordered domains. This phenomenon highlights the dynamic nature of thin-film formation during solvent evaporation, where polymer-solvent interactions and concentration gradients play key roles in determining the final morphology and optical properties of the film. The study of these processes provides valuable insight into the mechanisms of self-assembly and crystallization in polymer systems, which is crucial for optimizing thin-film fabrication techniques in various applications.

## 3.4 *Discussion*

The "polymer-solvent interactions" at play in the formation of concentration-dependent morphology in thin films are primarily governed by how the polymer chains interact with solvent molecules during the evaporation process. These interactions influence the behavior of the polymer as the solvent evaporates, leading to distinct morphologies in the resulting thin films. Polymer Solubility and Chain Conformation: At different concentrations, the polymer's solubility and the conformation of its chains in the solvent vary significantly. In dilute solutions, individual polymer chains are well-separated, allowing them to remain in an expanded or coil-like conformation. As the solvent evaporates, the polymer chains can aggregate more loosely, often resulting in less-ordered structures. In more concentrated solutions, the polymer chains are closer together, leading to stronger interactions between them. This proximity can promote the formation of more ordered, crystalline patterns, as the polymer chains align and crystallize in response to the shrinking solvent volume. Solvent Evaporation and Concentration Gradients: During evaporation, the solvent gradually leaves the system, leading to an increase in the local concentration of the polymer near the evaporating surface. This creates concentration gradients within the solution. In regions of higher concentration, the polymer chains may experience stronger interactions, leading to denser and more ordered structures. In contrast, regions with lower concentrations may exhibit more amorphous or loosely packed morphologies. Kinetics of Phase Separation: The rate of solvent evaporation also plays a critical role in determining the morphology. Faster evaporation rates can lead to kinetic trapping of non-equilibrium structures because the polymer chains do not have enough time to rearrange into more ordered states. Slower evaporation allows the polymer chains to reorganize and form more regular crystalline patterns. Therefore, concentration and evaporation rates together dictate the resulting morphology by influencing the kinetics of polymer-solvent phase separation. Intermolecular Forces: The nature of intermolecular forces, including van der Waals forces, hydrogen bonding, and dipole interactions, between the polymer and solvent molecules also affects the resulting thin-film morphology. In a good solvent, the polymer chains are well solvated, and the solvent molecules help keep the polymer chains dispersed. As the solvent evaporates, these interactions decrease, and the polymer chains start to collapse, aggregate, or crystallize, depending on the concentration. In a poor solvent or at high polymer concentrations, stronger polymer-polymer interactions dominate, leading to phase separation and crystallization. These polymer-solvent interactions, governed by factors such as polymer concentration, evaporation rate, and the nature of the solvent, lead to the observed concentration-dependent morphology in the thin films.

## 4. Conclusions

This study investigated the crystallization behavior of PAM solutions during evaporation and thin-film formation, with particular attention to the role of concentration gradients in the development of crystalline morphologies. As evaporation progresses, a concentration gradient emerges due to the uneven rate of solvent evaporation. In the outer regions where evaporation initiates more rapidly, the concentration remains relatively low, while in the central area, where the solvent persists longer, the concentration increases. This spatial variation in concentration leads to the formation of distinct crystalline structures. The findings revealed three main stages of crystallization, closely linked to the concentration profile across the evaporating droplet. Initially, at low concentrations, a third-level grass-like structure forms at the periphery. As evaporation continues, and the concentration increases in intermediate regions, second-level ridge-like or mountainous vein patterns emerge. Finally, in the central region, the highest concentration results in irregular, complex crystalline morphologies. The results from this study highlight the critical role of concentration gradients in determining the resulting crystallization patterns in PAM solutions. By providing detailed insights into how polymer concentration evolves during evaporation and influences the morphology of crystalline structures, this research advances our understanding of polymer crystallization mechanisms. Furthermore, the study's findings hold broader significance for various applications, such as the design of functional thin films in materials science. These insights pave the way for future work exploring how controlled evaporation and concentration modulation can be used to tailor crystalline structures for specific technological applications.

## Acknowledgment

This work was supported by the National Natural Science Foundation of China (No. 12202461), Shenzhen Science and Technology Research Program (Grant JCYJ20210324101610028)

## Conflict of Interest

The authors declare no competing financial interest.

[50]  Leong, K. H. Morphological control of particles generated from the evaporation of solution droplets: theoretical considerations. *J. Aerosol Sci.* **1987**, *18*, 511-524.

[51]  Halperin, A.; Tirrell, M. and Lodge, T. P. Tethered chains in polymer microstructures. Macromolecules: Synthesis, Order and Advanced Properties, **1992**, 31-71.19